# Involvement of calcium channels Orai3 in the chemoresistance to cisplatin in non-small cell lung cancer (NSCLC)

Daoudi Rédoane*

*E-mail : redoane.daoudi@unicaen.fr – University of Caen Normandy 14 000 Caen France

* Personal email address : red.daoudi@laposte.net

**Abstract.**

**This is the second part of the previous review. In the previous review we suspected that Orai3 channels were involved in lung cancer and more precisely in several cancers. Here we confirm that calcium dysregulation is important for cancer development. in this paper we show that Orai3 is an upstream activator of AKT and we prove that AKT is involved in chemoresistance in NSCLC.**

## Introduction

In 2012 in France, lung cancer was the 4th cancer in terms of incidence but the 1st in terms of mortality with a 5-year survival rate that does not exceed 5% in both men and women. This low survival rate is illustrated by the number of deaths attributable to this cancer alone each year in the world, indeed lung cancer is responsible for 1.6 million deaths per year on an international scale.

The main clinical sign which remains associated with lung cancer is of course respiratory failure characterized by difficult breathing which may be responsible for hypoxemia and secondary disorders. The alteration of the respiratory capacities can be visible by a decrease in FEV (Maximum Expiratory Volume Second). It should be noted that this clinical sign alone is not sufficient to make a diagnosis of lung cancer since a change in FEV1 can have several causes. The other clinical signs are noticeable weight loss, chest pain and cough possibly accompanied by blood spitting when intratumoral pro-angiogenic phenomena are important. It is important to note that some of these clinical manifestations are not due to the disease itself but rather to the treatments and therefore have an iatrogenic origin. In addition, not all patients have all the symptoms and their intensity depends on several factors. The stage of the disease, the constitution of the person, his age, the existence of possible other affections, the antecedents, the taking of other drugs are all elements modulating the progress of the disease for a given patient.

The most common cause of lung cancer is prolonged exposure to tobacco smoke, including passive smoking. Forms of cancer are also found in non-smokers and are thought to be due to a complex association of various environmental factors such as pollution and/or endogenous factors such as genetic predispositions.

Although the cellular context related to the tumor environment as well as tumor progression present interindividual variations, it is possible to identify common criteria on the basis of histological studies so that lung cancer can be classified into two major category. Thus 85% of lung cancers are so-called non-small cell cancers. The latter are subdivided into three subgroups: pulmonary adenocarcinoma (40% of cases), squamous cell carcinoma (40% of cases) and large cell carcinoma (20% of cases). The remaining 15% are small cell cancers.

The therapeutic management of lung cancer is mainly based on radiotherapy, surgery and chemotherapy. Concerning pulmonary adenocarcinomas, the first line of chemotherapeutic treatment is based on the association pemetrexed (Alimta®) and cisplatin. Cisplatin, a platinum salt, binds to the purine bases of the DNA (DeoxyriboNucleic Acid) of the cells, modifying the conformation of the DNA and thus inhibiting the phenomena of replication and transcription, which leads to the arrest of cell proliferation. Besides its cytostatic action, cisplatin also exhibits cytotoxic action.

If factors can modulate the severity of the disease for a given patient as seen above, there are also phenomena impacting the patient's response to treatment. Among these, it is well established that resistance mechanisms are set up following treatment and are responsible for a reduction in its effectiveness, with a poor prognosis: this is called acquired chemoresistance. Other mechanisms are present even before the administration of the treatment, these are innate chemoresistance. Chemoresistance results from complex changes in cell function. Among the different molecular actors studied, many clinical studies have highlighted the role of certain proteins of the family of ABC-type transporters (Atp Binding Cassette),Guminski et al., 2006). These proteins include a member of the MRP family (Multidrug Resistance Proteins) called MRP5, product of the ABCC5 gene (Atp Binding Cassette subfamily C member 5). The latter has been shown to be involved in resistance to an anti-cancer drug, gemcitabine, in pancreatic cancer (Wolfgang et al., 2011). In hepatocellular carcinoma, the expression of MRP5 has been correlated with resistance to cisplatin (Wakamatsu et al., 2007). Numerous other studies have established a correlation between the expression of MRP5 and resistance to multiple anticancer agents. The MRP5 protein and more generally MRP-type proteins therefore have the ability to cause the efflux of a broad spectrum of anti-cancer agents including cisplatin, thus causing a reduction in the efficacy of the cytotoxic agent.

Numerous signaling pathways allowing the regulation of the expression of MRP5 are thus involved in the phenomenon of chemoresistance and have been partly elucidated in the scientific literature. Among them, the AKT pathway would play a preponderant role.

Class I PI3K (Class I PhosphoInositide-3-Kinase) produces, among other things, PIP3 (PhosphatidylInositol(3,4,5)-triPhosphate) through the phosphorylation of a precursor PIP2 (PhosphatidylInositol(4,5)-diPhosphate). The PIP3 thus formed is a constituent of the plasma membrane and serves as a signaling lipid to activate the AKT pathway further downstream. Note also that PIP3 can be dephosphorylated into PIP2 for signal extinction, this dephosphorylation is mediated by the phosphatase

PTEN (Phosphatase and TENsing homolog) whose gene expression is frequently deregulated during the cancer process.

The PH domain (Pleckstrin Homology) is a protein domain made up of approximately 120 amino acids found in several proteins including AKT and PDK1 (Phosphoinositide-Dependent Kinase-1). This domain can bind PIP3 to the plasma membrane, creating a microenvironment consisting of AKT and PDK1. PDK1 then activates AKT by phosphorylating it on its threonine 308 (Michael et al., 1999). Activation is completed by PDK2 (Phosphoinositide-Dependent Kinase-2) which phosphorylates AKT on its serine 473 (Michael et al., 1999). The activation of AKT regulates many targets further downstream, which are involved in as many physiological or pathophysiological cellular processes as the regulation of glucose metabolism following the release of insulin, cell proliferation, cell survival, protein synthesis via the mTORC1 complex (Mammalian Target Of Rapamycin Complex 1), angiogenesis, tumorigenesis. During this last AKT was shown to be involved in chemoresistance to several anti-cancer compounds including cisplatin and this in several cancers (Fraser et al., 2008) (Abolfazl et al., 2016) (Zhang et al. , 2016). Among the many targets of AKT is GSK3β (Glycogen Synthase Kinase-3 β)which undergoes inhibitory phosphorylation by AKT (Dexing et al., 2007). In turn GSK3β has been shown to inactivate the C-MYC protein (Gregory et al., 2003), a product of the C-MYC proto-oncogene. However, several studies show the non-negligible role of C-MYC in the expression of MRP5. It has indeed been shown a possible positive regulation of MRP5 by C-MYC since the functional invalidation of C-MYC by an siRNA (Small Interfering RiboNucleic Acid) C-MYC is coupled with a decrease in the expression of the ABCC5 gene in the COLO-320 colorectal cancer cell line (Naruji et al., 2015). Moreover, numerous intracellular signaling pathways which depend on AKT and which lead to the expression of C-MYC could be involved in resistance to cisplatin via MRP5. This is the case of the β-CATENIN or SONIC-HEDGEHOG pathways for example.

A second important gene, this time a tumor suppressor, which is P53, may be linked to MRP5-dependent chemoresistance in cancer. In fact, it has been established that the non-mutated form of the P53 protein decreases the expression of the genes encoding the proteins of the MRP family, including MRP5 (Wang and Beck, 1998). Mutated forms of P53 decrease the expression of genes coding for proteins of the MRP family, but to a lesser extent. Unmutated and mutated forms of P53 would therefore be associated with sensitivity to cisplatin and other chemotherapeutic compounds. The link with AKT is obvious, it is well known that AKT activates MDM2 (Murine Double Minute 2), an E3 ubiquitin ligase of P53, by phosphorylating it on its serine 186 (Lindsey and David, 2001). AKT therefore increases the degradation of P53 and would induce chemoresistance in cell lines expressing the non-mutated form of P53, the expression of MRP5 being subsequently increased. Another signaling pathway potentially links AKT and P53 in our lung adenocarcinoma H23 cell line of interest. It has indeed been shown that AKT activates mTORC1 by inhibiting TSC2 (Tuberous Sclerosis Complex 2) (Bhaskar and Hay, 2007). In addition, mTORC1 activates the transcription factor HSF1 (Heat Shock Factor 1) in HeLa cells (Henrietta Lacks) (Shiuh-Dih et al., 2012). In turn, HSF1 activates the chaperone protein HSP27 (Heat Shock Protein 27) in non-small cell lung cancer A549 cells (Lelj-Garolla et al., 2015). HSP27 can activate MDM2 in HEK293A cells (Human Embryonic Kidney cells 293) (Xu et al., 2013) and would therefore be involved in resistance to cisplatin by causing the degradation of P53 and the increase in the expression of MRP5. All of these data place AKT as a major player in resistance and as a decision-making crossroads in the cell with the downstream modulation of two targets, C-MYC and P53, which could be linked to the modulation of MRP5, one of the players resistance to cisplatin.

Moreover, calcium is a very important cation in several physiological processes and in particular cell proliferation. AKT is also well known for its proliferative role, so a link between calcium signaling and AKT might exist. We therefore wished to study calcium signaling since such a link would justify the involvement of calcium in the processes of proliferation and especially of chemoresistance.

In cells, the intracellular calcium concentration is finely regulated. The players in calcium homeostasis are the membrane calcium channels whose mode of opening is varied, the ATP (Adenosine TriPhosphate)-dependent calcium pumps (PMCA (Plasma Membrane Calcium ATPase) at the plasma membrane, SERCA (Sarco/Endoplasmic Reticulum Calcium ATPase) to the membrane of the endoplasmic reticulum) as well as the exchangers (NCX (Sodium-Calcium Exchanger) to the plasma membrane). This concentration is of the order of 1.8 mmol/L in the extracellular compartment whereas it is only 0.0001 mmol/L in the intracellular compartment, thus providing an electrochemical gradient favorable to the entry of extracellular calcium towards inside the cell. This calcium enters cells through specific proteins called calcium channels in the plasma membrane. Since we want to study the impact of calcium in chemoresistance we focused our study on a particular calcium channel called ORAI3. The family of ORAI channels contains three known isoforms which are ORAI1, ORAI2 and ORAI3. These are channels activated by a depletion of intracellular calcium stores, in particular those of the endoplasmic reticulum. These channels are thus qualified as SOC (Store Operated Channels) and allow an entry of calcium called capacitive calcium entry or SOCE (Store Operated Calcium Entry). When the reticular calcium concentration decreases, a reticulum-resident calcium level sensor protein called STIM1 (Stromal Interaction Molecule 1) detects this decrease in calcium concentration and oligomerizes, is relocated to the plasma membrane and opens the ORAI3 channels. The subsequent calcium entry activates a specific task in the cell and replenishes the calcium stores in the lumen of the reticulum, essential for various functions such as protein folding in the reticulum.

We wanted to study the ORAI3 channel precisely since it was shown, within the laboratory, that the functional invalidation of the channel by means of siRNA was correlated with a decrease in the phosphorylation levels of AKT in H23 and H460 cells (Anne -Sophie et al., 2013) and therefore possibly a decrease in resistance via AKT and MRP5. In addition, in the MCF7 (Michigan Cancer Foundation-7) breast cancer lines, it was shown that the siRNA ORAI3 reduced the expression as well as the activity of C-MYC, confirming this idea (Faouzi et al., 2013) . Preliminary results from the lab also showed that siRNA ORAI3 was associated with a decrease in protein C-MYC in A549 cells this time around. These two results show that two a priori different lineages retain a common pathway involving ORAI3 and C-MYC, suggesting a selection of this pathway conferring a possible selective advantage to tumors. Other preliminary results from the laboratory showed an increase in the expression of the ORAI3 channel by cisplatin in A549 cells, thus demonstrating the existence of a possible acquired chemoresistance induced by cisplatin itself.

Pulmonary adenocarcinomas are at the heart of therapeutic concerns since their incidence continues to increase. The low 5-year survival rate mentioned initially can be partly explained by drug resistance.
All of the previous data then provides solid bases showing a probable involvement of the ORAI3 channel in cisplatin chemoresistance mediated by MRP5.

**Materials and methods**

***Culture experiment.*** The cell culture experiments were carried out in a category 2 culture room, under a type II biological safety hood, BH 2004 D, with vertical laminar flow.

***Cell culture medium.*** A 10% FCS (Fetal Calf Serum) solution was used as culture medium. A -SVF solution (without SVF) was prepared on the basis of a volume of 500 mL and was composed of: 385.4 mL of sterile water, 50 mL of EMEM 10X (Eagle's Minimum Essential Medium), 14.6 mL of bicarbonates 7.5%, 10mL HEPES 1M(4-(2-HydroxyEthyl)-1-PiperazineEthaneSulfonic acid),5mL of non-essential amino acids (MEM 100X (Minimum Essential Medium)), 5mL of gentamicin, 5mL of L-alanyl-L-glutamine 200mM (GlutaMax™, a more stable analog of L-glutamine providing L-alanine and L-glutamine). In order to obtain a 10% FCS solution, 45mL of -SVF solution was added to a tube and then topped up with 5mL of pure FCS.

***Cell preparation.*** The H23 cell line was used. These are cancerous cells of human alveolar epithelium (adenocarcinoma). These cells have mutations in the KRAS gene(*V-Ki-ras2 Kirsten rat sarcoma viral oncogene homolog) (G12C)*, which makes the RAS protein constitutively hyperactive by default of hydrolysis of GTP (Guanosine TriPhosphate) into GDP (Guanosine DiPhosphate) with exaggerated activation of the MAPK (Mitogen-Activated Protein Kinases) and PI3K/AKT pathways.

The H23 cells have a mutation of codon 246 of the P53 gene as well as an amplification of the proto-oncogene C-MYC (20 times compared to a non-cancerous cell), however without modification of the mRNA level (Messenger RiboNucleic Acid) of C-MYC.

The ampoules containing the cells were maintained in liquid nitrogen (-80℃) then thawed in a water bath at 37℃. The cells were then suspended in 9 mL of culture medium and then centrifuged (1000 revolutions/min for 10 minutes).

After centrifugation and to eliminate the freezing medium, the supernatant obtained was eliminated and the cells were resuspended in 1 mL of culture medium.

The suspension was placed in a 25cm flask containing 5mL of culture medium. This flask was then kept in an incubator containing 5% of $CO_2$(Carbon dioxide), an atmosphere saturated with humidity and at 37℃. After 24 hours of incubation, the cells adhere to the bottom of the flask and proliferate, forming a cell carpet. The medium was then renewed every 48 hours to optimize cell proliferation.

After 6 to 7 days of culture, the cells occupy 80% of the surface of the flask, justifying a detachment of the cells by means of a treatment with trypsin to avoid inhibition of proliferation by contact and thus avoid distorting the results of the subsequent experiences.

***Detachment of cells.***The culture medium was removed using an aspiration syringe then 10 mL of PBS (Phosphate Buffered Saline) were added in order to eliminate the trypsin inhibitors present in the culture medium previously removed. The PBS was then aspirated.

400μLof trypsin were then applied to the cells, together with a chelator of divalent cations, EDTA (Ethylene Diamine Tetra-Acetic acid). This step made it possible to detach the cells from the culture dish. After a few minutes of waiting, the trypsin previously applied was neutralized by adding 10 mL of culture medium, reflux was carried out to homogenize the cell suspension.

***Malassez cell count.***Cell counts were performed on Malassez cell.

***Transient transfections.***A detachment of the cells was initially carried out according to the protocol described above.Afterwards2 eppendorfs of 1.5mL (1 eppendorf SiRNA Ctrl, 1 eppendorf SiRNA ORAI3) each containing$10^6$cells were prepared.

The eppendorfs were then centrifuged (1000 rpm for 7 minutes, 4℃). The supernatant was aspirated and the pellet containing the cells was saved.

100μLofnucleovector®Kit V were added in the eppendorfs to resuspend the pellet then2μg of SiRNA Ctrl and SiRNA ORAI3, ie 10 μL of SiRNA Ctrl and SiRNA ORAI3, were each added in a different eppendorf. All of the medium contained in the eppendorfs was taken and then placed in a transfection tank (1 SiRNA Ctrl tank, 1 SiRNA ORAI3 tank).

Transfection was then carried out using the Nucleofector II transfection apparatus (transfection by electroporation) set up with the appropriate transfection program. (program MDA-MB-231, X013).500μLof culture medium were added to each of the two tanks to restore the cells to favorable conditions, the medium present in the two tanks was then removed and then transferred back to two separate eppendorfs. The two eppendorfs were kept at 37°C for 5 minutes before transferring the transfected cells to the boxes provided for the various tests of the study, these boxes were placed in an incubator until they were used. All the transfections were validated using the RT-PCR (Reverse Transcription Polymerase Chain Reaction) method described below.

***MTT assay.***The MTT test(3-(4,5-dimethylthiazol-2-yl)-2,5-diphenyl tetrazolium bromide) was used to estimate cell viability. MTT is metabolized by succinate dehydrogenase in the mitochondria of living cells to form formazan which forms a purple precipitate, visible at 550 nm. The reading of the absorbance at this wavelength thus provides information on the rate of viable cells between several conditions.

The cells were introduced into each well provided for the MTT test with 1.5 mL of culture medium in each well. The next day the cells adhered to the bottom of the well and the culture medium was aspirated. A volume of 1.5 mL of medium containing cisplatin at varying concentrations was added to the wells. For the well without cisplatin, 1.5 mL of culture medium alone was added. The incubation of the MTT wells with cisplatin lasted 48 h. 48 hours later 800µLof culture medium containing 0.5 g/L of MTT was added to each of the wells. The MTT was stored in the dark and in the refrigerator. The wells were then incubated for 60 minutes in the presence of MTT and then the medium containing the MTT was removed by aspiration.

1000µLof DMSO (DiMethylSulfOxide) were added to each well to recover the formazan formed by the living cells then the medium containing the formazan was transferred to a 96-well plate before reading the absorbance at 550nm taking into account the background noise of the plaque. During the transfer, each well of the 96-well plate was filled with 200 µL of formazan present in the wells of the MTT plate.

The Nanodrop 2000 spectrophotometer was used for the automated reading of the absorbances in each well, software attached to the device made it possible to present the results in an Excel file.

***Calcium imaging: capacitive calcium input.***Intracellular calcium concentrations were estimated with the fura-2 probe and Metafluor software. The H23 cells were incubated at 37° C. in the presence of the fura-2 probe for 45 minutes. The probe was then excited at 340nm in the calcium-bound form and at 380nm in the free, non-calcium-bound form. Fluorescence intensity was measured at 512nm. The ratiofluorescence intensities measured during excitations at 340nm and 380nm, denoted R, was determinedunder Metafluor. We pose $R = \frac{IF340}{IF380}$ with IF the fluorescence intensity measured at 340nm or 380nm. The more R increases, the higher the calcium concentration. In this study 2mM calcium was infused into the extracellular medium for 60 seconds. From 60 seconds to 120 seconds the 2mM calcium infusion was turned off and calcium-free medium was infused instead to remove extracellular calcium. At 2 minutes, a calcium-free medium containing 1 µM thapsigargin was perfused into the extracellular medium. Thapsigargin is an inhibitor of the SERCA pumps of the endoplasmic reticulum which actively transport calcium to the lumen of the reticulum. This thus allows the emptying of the calcium stocks of the reticulum and the exit of the calcium towards the cytoplasm. This emptying allows the activation of SOC channels like ORAI3 at the plasma membrane. After infusion the ratio increases due to the presence of calcium in the cytoplasm in greater quantity. Subsequently, the intervention of various players in calcium homeostasis (PMCA pumps, NCX exchangers) is responsible for a reduction in the ratio. The thapsigargin infusion was closed when the ratio began to decrease, while the calcium-free one was opened. Around 15 minutes, the ratio R becomes minimal and Rmin is set as being this ratio. At 15 minutes 10mM of calcium was infused into the extracellular medium, responsible for an increase in the ratio, the calcium entering the cells through the previously activated SOC channels. The ratio becomes maximum and Rmax is noted as being the maximum ratio obtained after addition of 10 mM of calcium. This ratio then decreases thanks to the actors of calcium homeostasis described above. We define $\Delta R = Rmax - Rmin$ And $\dfrac{\Delta R}{Rmin} = \dfrac{Rmax - Rmin}{Rmin}$. More the report $\dfrac{\Delta R}{Rmin}$ is greater the greater the capacitive calcium influx induced by thapsigargin. During the experiment, the ratios were automatically saved in the Origin software.

***Real-time quantitative RT-PCR.*** The cell medium was aspirated then 1mL of Trizol® was added to lyse the cells. The lysate was then stored at -80°C until extraction of the nucleic acids. On the day of extraction, the cell lysate was thawed then the total RNAs were extracted.

After drying, the pellet was rehydrated with 22μL of water. The amount of genetic material was estimated by measuring absorbance with the Nanodrop spectrophotometer. For that, 1μL samples were introduced at the level of the spectrophotometer and a concentration of total RNAs in the solution as well as the purity of the total RNAs were obtained. A reverse transcription from 2μg of total RNAs was then carried out.

The purpose of this step is to synthesize cDNA (complementary DNA) of RNA in order to allow the polymerase used in qPCR (Quantitative Polymerase Chain Reaction) to bind to its substrate. To solutions containing 2μg of RNA, or 0,2μg/μL of RNA, were added 10μL of reaction mix. For 10μL, the mix contained 1μL RT (Reverse Transcriptase), 1μL RNase inhibitors, 0,8μL dNTP (DeoxyriboNucleotides TriPhosphates) (100nM), 2μL primers (Reverse and Forward), 2μL of RT buffer and 3,2μL sterile water (experimental data provided with the reverse transcription kit). The reverse transcription cycle started with 10 minutes at 25°C in order to prepare the samples. Then, the samples were brought to 35°C for 120 minutes, which allowed the hybridization of the primers and the synthesis of the cDNAs. Finally the samples were brought to 85°C for 5 minutes in order to stop the synthesis and separate the primers.

Quantitative PCR reactions were then performed on the LightCycler® from Roche in 420-well microplates and using the ABsolut kitTMqPCR SYBR® Green Mixes (ABgene) in accordance with the supplier's instructions. To do this, a DNA intercalator, SYBR green, was used, which emits fluorescence in contact with double-stranded DNA, with a maximum intensity at 550 nm. This fluorescence was detected and measured by the LightCycler microspectrofluorimeter®, which allowed quantification of the number of copies of the cDNA sequence of interest. At each end of the cycle, the fluorescence is therefore measured and it is possible to visualize the exponential increase in real time of the quantity of amplicons generated. Finally, the number of cycles necessary for the fluorescence of the target gene to be detectable is inversely proportional to the initial quantity of the mRNA of interest. This number of cycles was calculated and compared for a given gene between different conditions. The target sequences were amplified from specific primers and using polymerase activating at high temperature. For each sample, a reaction medium was prepared containing 4,6μL of reaction mix (SYBR®Green, Taq (Thermus AQuaticus) polymerase, dNTP, enzyme buffer and MgCl2), 2,2μL of cDNA diluted to the 20th, 0,5μL to prime forward, 0,5μL primer reverse (both at 102μM)

And 1,6μL sterile water.

The reference gene used was actin, in order to verify that the variations in the number of cycles obtained for a gene of interest between different conditions were indeed due to a difference in the initial quantity of mRNAs of this gene and not to the presence of Taq polymerase inhibitors between two conditions or the fact that less genetic material was collected between two conditions.

***Statistics.*** The results of the different experiments were analyzed using a two-way ANOVA (ANalysis Of Variance) test. The significance threshold being $p < 0.05$. The software used for the statistics was SigmaStat 3.0.

**Results**

**1. Involvement of the ORAI3 calcium channel in cisplatin resistance in H23 cells.**

We first wanted to study the effect of cisplatin on our H23 cell line after 48 hours of incubation at increasing concentrations of the molecule. For this, the cell viability of the H23 cells was evaluated using an MTT test. This pharmacological approach made it possible to draw an effect/dose curve and to determine the concentration of cisplatin for which the viability is reduced by half. This concentration is called IC50 (the half maximal Inhibitory Concentration). The parameter observed here is the absorbance (optical density) read at 550 nm, reflecting the amount of formazan formed by living cells and therefore the proportion of living cells as a function of cisplatin concentrations (viability).

We saw in the introduction of this thesis that the calcium channel ORAI3 could be involved in resistance to cisplatin in H23 cells. For this reason we also wanted to know if this channel actually played a role in the resistance to cisplatin in this line. For this we performed the same MTT test by functionally invalidating the ORAI3 channel using the interfering RNA technique (siRNA ORAI3) and we determined the corresponding IC50. The IC50 was compared to the population of cells whose ORAI3 channel was not knocked out (siRNA Ctrl).

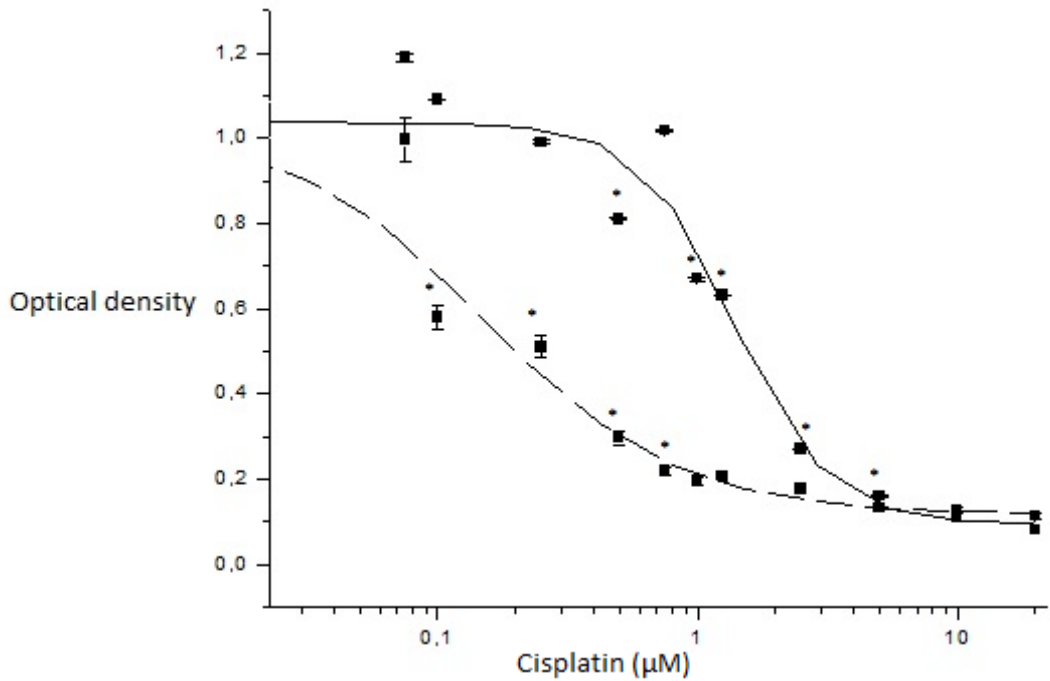

Figure 1: determination of the cell viability of H23 cells whose ORAI3 channel has been invalidated or not, 72 hours post-transfection (n=1). Incubation in cisplatin for 48 hours. Seeding of 70,000 cells per well of the MTT plate. A cell population without transfection (other than siRNA Ctrl) was also studied and gave an IC50 value of approximately 2 µM (not shown here, n=5). The doses of cisplatin used were as follows: 0 µM – 0.075 µM – 0.1 µM – 0.25 µM – 0.5 µM – 0.75 µM – 1 µM – 1.25 µM – 2.5 µM – 5 µM – 10 µM – 20 µM. ·········· siRNA ORAI3 ——— siRNA Ctrl * significant result (p < 0.05).

Cisplatin induced a decrease in the cell viability of H23 in a dose-dependent manner in the two cell populations (siRNA ORAI3 and siRNA Ctrl).For low concentrations of cisplatin, the cell viability remains unchanged in the two cell populations (up to 0.5 µM in the cells with siRNA Ctrl and below 0.01 µM for the cells with siRNA ORAI3). Then the viability decreases then stabilizes (plateau from 10 µM in the cells with siRNA Ctrl and 5 µM for the cells with siRNA ORAI3).In cells where the ORAI3 channel has not been invalidated, the IC50 is 2 µM.This concentration was used for all subsequent experiments.The cells which have been transfected with the siRNA ORAI3 show an IC50 which is significantly reduced compared with the cells transfected with the siRNA Ctrl. This is equal to approximately 0.25 µM of cisplatin, ie a reduction of almost 8 times compared to the IC50 of cells whose ORAI3 channel has not been functionally invalidated.

In parallel, the transfection was verified using the RT-PCR technique. ORAI3 channel mRNA levels were halved (not shown here).

**2. Impact of the functional invalidation of the ORAI3 channel on the capacitive entry of calcium into H23 cells.**

For the rest of our study, we wanted to explain at the intracellular level the phenotype observed in the MTT test for cells whose ORAI3 channel has been invalidated, that is to say the reduction in IC50. More specifically, we wanted to study the impact of the absence of the ORAI3 channel on calcium homeostasis, perhaps in connection with the modification of IC50 observed when the channel is invalidated. For this we used the calcium imaging technique and evaluated the capacitive calcium influx triggered by thapsigargin, an inhibitor of the SERCA pumps of the endoplasmic reticulum. The effect of cisplatin on capacitive calcium entry, in the presence and absence of the functional ORAI3 channel, was therefore determined in order to be able to relate the changes in capacitive calcium input observed with the change in IC50 observed in the MTT test. The cells were incubated for 72 hours in the presence of cisplatin at 2 µM.

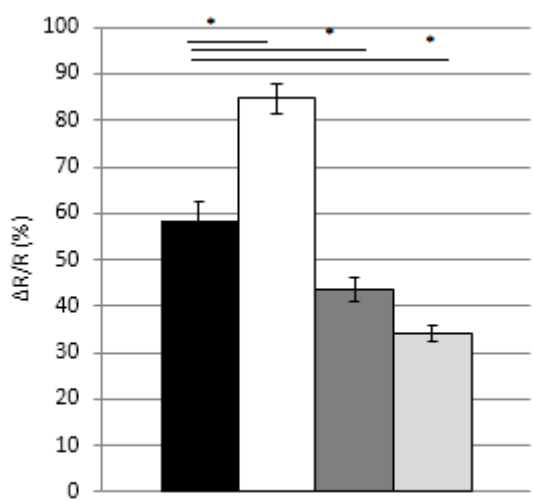

Figure 2: measurement of the capacitive entry of calcium into the H23 cells whose ORAI3 channel has been invalidated or not, in the presence and absence of cisplatin at 2 µM, 96 hours post-transfection (n=1). Incubation in cisplatin for 72 hours. ■ siRNA Ctrl □ siRNA Ctrl + 2 µM cisplatin ▩ siRNA ORAI3 □ siRNA ORAI3 + 2 µM cisplatin * significant result (p < 0.05).

The ratio $\dfrac{\Delta R}{Rmin}$ , reflecting the capacitive influx of calcium, is increased by 26% in the presence of 2 µM cisplatin compared to cells which have not been incubated with cisplatin. On the other hand, when the ORAI3 channel is functionally invalidated and still in the presence of cisplatin, the capacitive calcium entry is significantly reduced (approximately 25%) compared to cells that have not been incubated with cisplatin and whose ORAI3 channel is functional. The invalidation of the ORAI3 channel alone, without cisplatin, causes a decrease in capacitive calcium entry of almost 15% compared to cells not incubated with cisplatin but whose ORAI3 channel is functional.

**3. Impact of functional invalidation of the ORAI3 channel on the expression of target genes involved in cisplatin resistance in H23 cells.**

To continue to explain at the intracellular level the decrease in IC50 observed in the absence of the functional ORAI3 channel, we continued our study by analyzing the expression of genes coding for actors possibly involved in resistance to cisplatin in the H23 cells. Among them, the expression of the gene coding for the MRP5 protein, already described in the introduction as possibly being involved in resistance to cisplatin, was evaluated in the absence and presence of the functional ORAI3 channel. The effect of cisplatin on the expression of these genes, in the presence of the functional ORAI3 channel, was also evaluated.

The expression of four genes was evaluated using the real-time quantitative RT-PCR technique, in the presence and absence of the ORAI3 channel as well as in the presence of 2 μM of cisplatin.

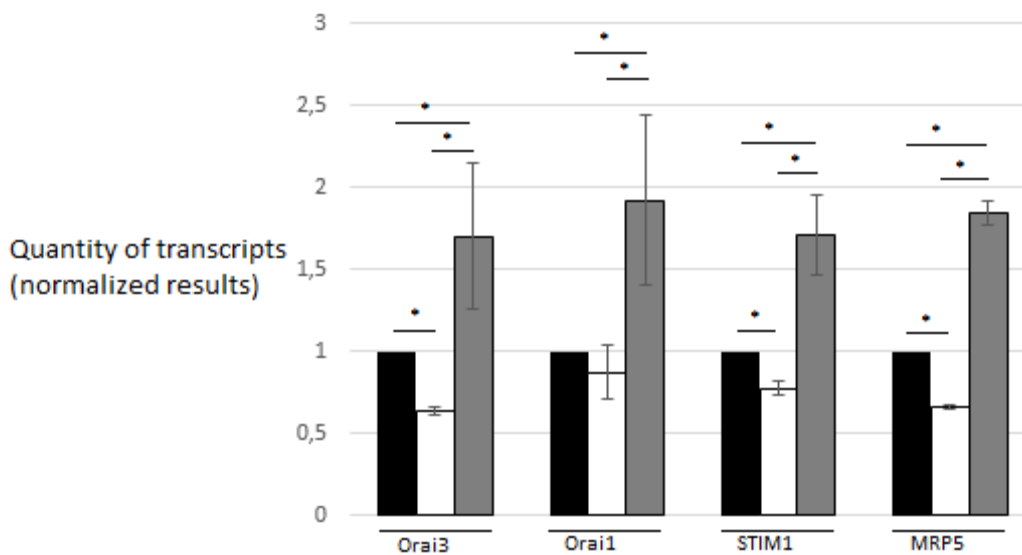

Figure 3: Gene expression analysis *ORAI3, ORAI1, STIM1 and ABCC5 (MRP5 protein) in H23 cells in the absence of the functional ORAI3 channel or in the presence of cisplatin* at 2 μM, 96 h post-transfection (n=1). Incubation in cisplatin for 72 hours. Results normalized to control. The reference gene used was actin. ■ siRNA Ctrl □ siRNA ORAI3 ▤ siRNA Ctrl + 2 μM cisplatin * significant result (p < 0.05).

Functional invalidation of the ORAI3 channel leads to a significant decrease in the mRNA level of the ORAI3 channel (decrease from 1 to 0.63, transfection validated). STIM1 and MRP5 mRNA levels are also decreased in the absence of the functional ORAI3 channel (1–0.77 for STIM1, 1–0.66 for MRP5). The observed decrease for ORAI1 mRNA is not significant (1 to 0.87).

In the presence of cisplatin at 2 µM and in the presence of the functional ORAI3 channel, the mRNA levels of the four genes of interest are significantly increased (1 to 1.70 for ORAI3, 1 to 1.92 for ORAI1, 1 to 1 .71 for STIM1, 1 to 1.84 for MRP5).

## 4. Influence of chronic exposure to cisplatin on the IC50.

We then incubated the H23 cells for 4 weeks in the presence of 2 µM cisplatin. Indeed, the incubation for 4 weeks is close to reality since a course of cisplatin lasts 4 weeks for patients with non-small cell lung cancer. We wanted to know whether this chronic exposure to cisplatin significantly modified the IC50 already determined for cells having undergone acute exposure to cisplatin (48 h).

In order to test a possible modification of the IC50 for the cells having been incubated chronically in the presence of cisplatin, we carried out an MTT test. Cisplatin concentrations were adjusted in anticipation of probable acquisition of cisplatin resistance conferred by chronic exposure to the product.

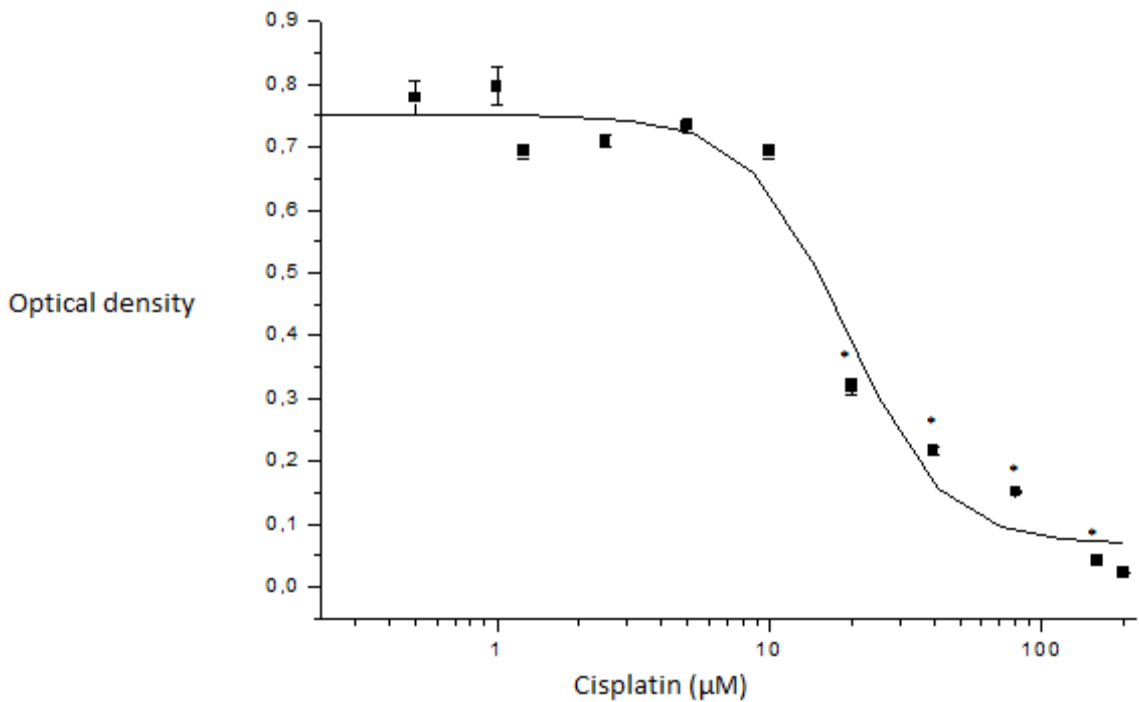

Figure 4: determination of cell viability of H23 cells incubated for 4 weeks in the presence of cisplatin at 2 µM (n=1). Seeding of 70,000 cells per well of the MTT plate. After 1 month of incubation, the doses of cisplatin used for the test were as follows: 0 µM – 0.5 µM – 1 µM – 1.25 µM – 2.5 µM – 5 µM – 10 µM – 20 µM – 40 µM – 80 µM – 160 µM – 200 µM. * significant result (p < 0.05).

Up to 10 µM of cisplatin cell viability is unchanged. The latter then decreases significantly up to 100 µM of cisplatin before becoming stable again after 100 µM of cisplatin. The IC50 determined is equal to 20 µM of cisplatin, ie an IC50 10 times greater than that determined for cells incubated acutely in cisplatin.

**5. Influence of chronic cisplatin exposure on the expression of target genes involved in cisplatin resistance in H23 cells.**

Finally, we wanted to know whether this increase in IC50 for cells chronically incubated in cisplatin could be due to a different expression profile of the genes involved in resistance to cisplatin, in particular ABCC5 (protein MRP5).

To verify the hypothesis of a modification in the expression of the genes involved in resistance to cisplatin, we carried out a quantitative RT-PCR in real time for the four genes previously studied.

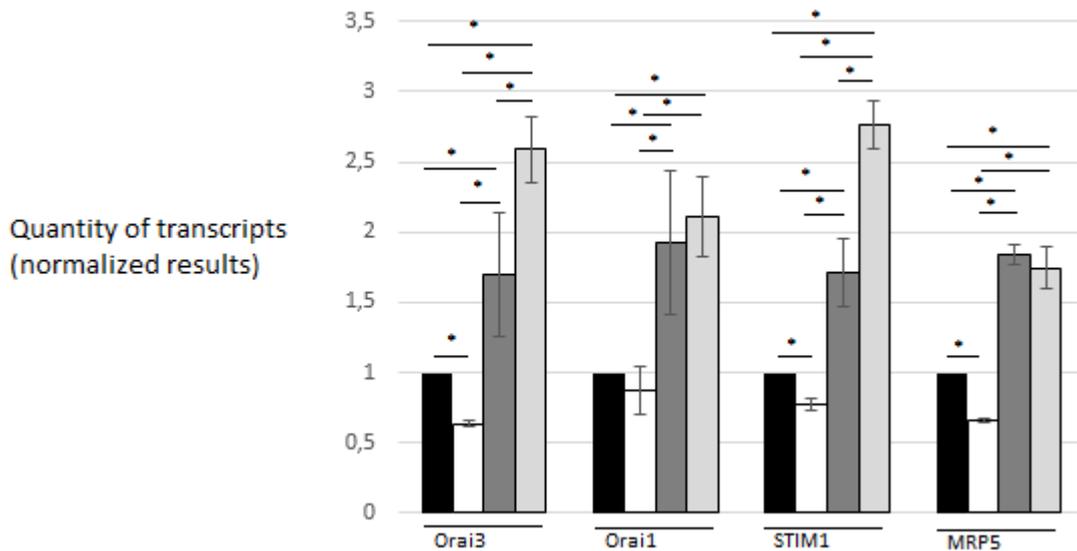

Figure 5: Gene expression analysis **ORAI3, ORAI1, STIM1 and ABCC5 (MRP5 protein) for H23 cells incubated for 4 weeks in the presence of cisplatin** at 2 μM (n=1). Results normalized with respect to the control. The reference gene used was actin. ■ siRNA Ctrl □ siRNA ORAI3 ▣ siRNA Ctrl + 2 μM cisplatin (72h incubation) □ siRNA Ctrl + 2 μM cisplatin (4 weeks incubation) * significant result (p < 0.05).

Chronic cisplatin incubation induced a significant increase in ORAI3 and STIM1 mRNA levels compared to acute incubation (1.70 to 2.58 for ORAI3, 1.71 to 2.76 for STIM1). The changes in the mRNA level of ORAI1 and MRP5 are not significant (1.92 to 2.11 for ORAI1, 1.84 to 1.74 for MRP5).

**Discussion**

For the whole of our study, we considered any increase in the IC50 as an increase in resistance to cisplatin for the H23 cells. The IC50 determined for cells acutely incubated in cisplatin, is 2 µM. This data agrees with the bibliography which shows values close to our IC50 for H23 cells. One study gives a value of 0.74 µM for the IC50 of cells acutely incubated in cisplatin (96h) (Lin et al., 2016). We hypothesize that the observed difference between our IC50 and that of the literature is due to different cell culture conditions. Even small differences may be responsible for the observed variation. In particular, the authors incubated the H23 cells in cisplatin for 96 hours before carrying out the MTT test and not 48 hours.

The functional invalidation of the ORAI3 channel is responsible for a decrease in the IC50 of nearly 8 times, bringing the value from 2 µM to 0.25 µM. In the MTT test and for the condition without cisplatin, the cell viability is unchanged between the cells whose ORAI3 channel is invalidated and the cells whose ORAI3 channel is functional. This indicates that the decrease in cell viability under the other conditions (in the presence of different doses of cisplatin) is not due to basal mortality which would be the consequence of the absence of the ORAI3 channel but rather to a greater sensitivity to cisplatin. In H23 cells and in the absence of the functional ORAI3 channel, resistance to cisplatin is therefore reduced since a lower concentration of cisplatin is required to induce a reduction in cell viability by half.

Moreover, the cells transfected with a Ctrl siRNA show an IC50 relatively close to that obtained with non-transfected cells (results not shown here). The minimal variation of the IC50 observed between the cells having integrated the siRNA Ctrl and the non-transfected cells may be due to the transfection. In fact, the electroporation step carried out during the transfection can temporarily modify the ion fluxes passing through the cell, including those passing through the ORAI3 channel, with finally small modifications of the IC50.

The increase in capacitive calcium influx induced by cisplatin could be due either to an increase in the expression of functional ORAI3 channels at the plasma membrane or to a modification in the type of response of the ORAI3 channels which would change from a state non-SOC to a SOC state. We favor the first hypothesis since, as announced in the introduction, the preliminary results of the laboratory have already shown that cisplatin was able to induce an increase in the mRNA level of the ORAI3 channel in A549 cells. This result could therefore also be observed in H23 cells. The increased functionality of the channel then remains to be confirmed. The functional invalidation of the ORAI3 channel coupled with the presence of cisplatin is associated with a decrease in the capacitive entry of calcium induced by cisplatin. This result could be explained by the fact that the increase in the mRNA level of the ORAI3 channel by cisplatin is counteracted by the ORAI3 siRNA. Finally, the level of functional ORAI3 channels would even be reduced compared to cells without cisplatin and whose ORAI3 channel has not been invalidated since the capacitive calcium input is even lower than the control condition. This result also indicates that cisplatin could increase the expression of ORAI3 channels functioning in an SOC state since the functional invalidation of the ORAI3 channel coupled with the presence of cisplatin lowers the capacitive calcium input compared to the case where cisplatin is present. but without invalidating the channel. Finally, the level of functional ORAI3 channels would even be reduced compared to cells without cisplatin and whose ORAI3 channel has not been invalidated since the capacitive calcium input is even lower than the control condition. This result also indicates that cisplatin could increase the expression of ORAI3 channels functioning in an SOC state since the functional invalidation of the ORAI3 channel coupled with the presence of cisplatin lowers the capacitive calcium input compared to the case where cisplatin is present. but without invalidating the channel. Finally, the level of functional ORAI3 channels would even be reduced compared to cells without cisplatin and whose ORAI3 channel has not been invalidated since the capacitive calcium input is even lower than the control condition. This result also indicates that cisplatin could increase the expression of ORAI3 channels functioning in an SOC state since the functional invalidation of the ORAI3 channel coupled with the presence of cisplatin lowers the capacitive calcium input compared to the case where cisplatin is present. but without invalidating the channel.

Moreover, the sole invalidation of the ORAI3 channel also causes a decrease in the capacitive entry of calcium compared to the control condition, indicating that the transfection seems to be validated at the functional level and that the ORAI3 channels are, at least in part, involved in this phenomenon of capacitive calcium entry and therefore function as SOC-type channels even in the absence of cisplatin.

The decrease in resistance observed in the absence of the functional ORAI3 channel is then correlated with the decrease in capacitive calcium entry and a link could therefore exist between these two phenomena.

The functional invalidation of the ORAI3 channel also causes a decrease in the mRNA level of STIM1 and MRP5 (in addition to ORAI3). This could be due to the decrease in capacitive calcium input resulting in less activation of signaling pathways linking the ORAI3 channel and the activation of the transcription of these two genes. In particular, we hypothesize that the PI3K/AKT pathway would be less activated in the absence of the functional ORAI3 channel since in the introduction we mentioned the fact that the functional invalidation of the ORAI3 channel was coupled with a decrease in phosphorylation of AKT in H23 cells (Anne-Sophie et al., 2013). In addition, the activity and expression of the C-MYC gene, one of the indirect targets of AKT, was reduced in MCF7 cells (Faouzi et al., 2013). Preliminary results from the laboratory also revealed a reduced protein quantity of C-MYC in A549 cells in the absence of the functional ORAI3 channel. These data support the hypothesis of less activation of the AKT pathway when the ORAI3 channel is disabled.

Since C-MYC seems to be involved in the expression of the ABCC5 gene, we hypothesize that the AKT pathway being less activated, the C-MYC gene is less expressed and the C-MYC protein less stable and the ABCC5 gene is less expressed. , which results in less resistance to cisplatin in H23 cells in the absence of the functional ORAI3 channel. In the same way, the P53 protein being involved in the repression of the expression of the genes encoding the MRPs (Wang and Beck, 1998) it is possible that the decrease in activation of the AKT pathway leads to an increase in the activity of P53 and therefore a lower expression of the genes encoding the MRPs, which again results in a lower resistance to cisplatin in the H23 cells in the absence of the functional ORAI3 channel. However, H23 cells possess a mutated form of the P53 protein,

A correlation has also been shown between the decrease in activity of the PI3K/AKT pathway and the expression of the STIM1 gene in prostate cancer cells (Selvaraj et al., 2016). The decrease in expression of the STIM1 gene observed in our study could also be the consequence of an adaptation of the cell to a reduced capacitive calcium input. Indeed, the STIM1 protein being one of the actors of this capacitive calcium input, together with the ORAI3 channel, if this input is reduced following the functional invalidation of the ORAI3 channel then its expression rate could be reduced. In all cases, the decrease in the expression of the STIM1 gene could also explain the decrease in capacitive calcium input.

Cisplatin incubation increases ORAI3, ORAI1, STIM1 and MRP5 channel mRNA levels. This result suggests that cisplatin could be involved in acquired resistance to itself since MRP5 plays a role in cisplatin efflux (Wakamatsu et al., 2007). It is likely that cisplatin increases the expression of functional ORAI3 channels and therefore causes greater capacitive calcium input. Subsequently, the AKT pathway could be activated more significantly with an increase in the level of transcripts of these four genes. Cisplatin could also increase the transcription of these genes independently of the capacitive calcium input mediated by the ORAI3 channel, by acting directly on these genes at the nuclear level or through other signaling pathways. Functional invalidation of the ORAI3 channel does not significantly decrease the mRNA level of the ORAI1 channel. This result suggests that the expression of the gene encoding this channel is not dependent on the presence of the functional ORAI3 channel. This result also shows that cisplatin increases the expression of the gene coding for the ORAI1 channel independently of ORAI3. Like ORAI3, the ORAI1 channel could activate signaling pathways related to the expression of the ABCC5 gene and therefore contribute to innate and acquired resistance. The cisplatin-induced increase in capacitive calcium influx could therefore also be attributed to the ORAI1 channel, in addition to ORAI3 and potentially other players. This result suggests that the expression of the gene encoding this channel is not dependent on the presence of the functional ORAI3 channel. This result also shows that cisplatin increases the expression of the gene coding for the ORAI1 channel independently of ORAI3. Like ORAI3, the ORAI1 channel could activate signaling pathways related to the expression of the ABCC5 gene and therefore contribute to innate and acquired resistance. The cisplatin-induced increase in capacitive calcium influx could therefore also be attributed to the ORAI1 channel, in addition to ORAI3 and potentially other players. This result suggests that the expression of the gene encoding this channel is not dependent on the presence of the functional ORAI3 channel. This result also shows that cisplatin increases the expression of the gene coding for the ORAI1 channel independently of ORAI3. Like ORAI3, the ORAI1 channel could activate signaling pathways related to the expression of the ABCC5 gene and therefore contribute to innate and acquired resistance. The cisplatin-induced increase in capacitive calcium influx could therefore also be attributed to the ORAI1 channel, in addition to ORAI3 and potentially other players. This result also shows that cisplatin increases the expression of the gene coding for the ORAI1 channel independently of ORAI3. Like ORAI3, the ORAI1 channel could activate signaling pathways related to the expression of the ABCC5 gene and therefore contribute to innate and acquired resistance. The cisplatin-induced increase in capacitive calcium influx could therefore also be attributed to the ORAI1 channel, in addition to ORAI3 and potentially other players. This result also shows that cisplatin increases the expression of the gene coding for the ORAI1 channel independently of ORAI3. Like ORAI3, the ORAI1 channel could activate signaling pathways related to the expression of the ABCC5 gene and therefore contribute to innate and acquired resistance. The cisplatin-induced increase in capacitive calcium influx could therefore also be attributed to the ORAI1 channel, in addition to ORAI3 and potentially other players.

Finally, we wanted to study the effects of prolonged exposure to cisplatin on the IC50, since such a protocol is close to reality insofar as a course of cisplatin lasts approximately one month for patients with lung cancer not with small cells. The IC50 for cells that have been chronically incubated in cisplatin is 20 μM. Cells which are previously incubated for one month in the presence of cisplatin therefore develop greater resistance than cells not incubated in cisplatin and simply placed in the presence of the product for 48 hours thereafter. This increase in resistance may perhaps go through a modification of the level of transcripts of the actors involved in the resistance to cisplatin. For this reason, a real-time quantitative RT-PCR was performed on this so-called resistant line. The mRNA levels of ORAI3 and STIM1 are significantly increased compared to cells incubated for 48 h in cisplatin but not those of ORAI1 and MRP5. This can be explained in at least two ways. Cisplatin could cause a greater increase in the mRNA level of ORAI1 and MRP5 compared to cells acutely incubated in cisplatin. However, this greater increase could be transient only, the mRNA levels of ORAI1 and MRP5 being subsequently similar to those obtained in acute incubation. Nevertheless, the chemoresistance developed could still be greater than that developed in acute incubation since this greater expression, even transient, could be sufficient to establish greater drug resistance for several weeks thereafter. For example, a greater expression of the ABCC5 gene could translate into a greater quantity of functional MRP5 proteins. Then the level of mRNA expression of MRP5 would decrease again but the protein levels of MRP5 would remain constantly higher than those observed in the cells incubated acutely in cisplatin. It could therefore be an adaptation for the cell to a situation during which the level of functional MRP5 was transiently very high. This initial situation could have been a signal for the cell allowing it to subsequently maintain high protein levels of MRP5, even though MRP5 mRNA levels decreased again to become similar to those observed for acutely incubated cells. For example, an adaptation of the ubiquitin/proteasome degradation system could exist, with less degradation of MRP5 proteins.

The second explanation is that this increased expression is cyclical and no longer transient, and that the quantitative real-time RT-PCR was performed at a time when the mRNA levels of ORAI1 and MRP5 returned to similar to those observed for the cells in acute incubation.

In both of these scenarios functional MRP5 protein levels for chronically incubated cells could be higher than those of MRP5 for acutely incubated cells, even though the increase in MRP5 mRNA level during chronic incubation is not linear over the entire duration of the treatment, i.e. one month. This would then explain the greater cisplatin resistance obtained for chronically incubated cells compared to acutely incubated cells.

It remains to be determined how capacitive calcium entry occurs in H23 cells. Indeed, we have advanced the hypothesis that a greater proportion of functional ORAI3 channels is associated with greater capacitive calcium input and therefore greater resistance via greater activation of the AKT pathway or other pathways. Capacitive calcium entry would therefore play a determining role in the chemoresistance phenotype in H23 cells. However, it is known that this entry requires an emptying of the calcium stocks of the lumen of the endoplasmic reticulum. Recently it was shown that A549 cells express different types of muscarinic receptors (Zhao et al., 2015) and that they are able to secrete acetylcholine (Xu et al., 2013). M3 (Muscarinic) receptors, coupled to a Gq protein, could therefore be stimulated in an autocrine manner by the acetylcholine released and lead to the production of IP3 (Inositol Triphosphate) stimulating the IP3 receptor of the endoplasmic reticulum. The exit of calcium from the lumen of the reticulum towards the cytoplasm would then allow the emptying of the calcium stores of the reticulum and the capacitive entry of calcium by the ORAI3 channels of the plasma membrane. Such a mechanism has not been shown in H23 cells but could therefore exist, given the proximity of cell types between A549 and H23 cells. The exit of calcium from the lumen of the reticulum towards the cytoplasm would then allow the emptying of the calcium stores of the reticulum and the capacitive entry of calcium by the ORAI3 channels of the plasma membrane. Such a mechanism has not been shown in H23 cells but could therefore exist, given the proximity of cell types between A549 and H23 cells. The exit of calcium from the lumen of the reticulum towards the cytoplasm would then allow the emptying of the calcium stores of the reticulum and the capacitive entry of calcium by the ORAI3 channels of the plasma membrane. Such a mechanism has not been shown in H23 cells but could therefore exist, given the proximity of cell types between A549 and H23 cells.

A second mechanism independent of the emptying of calcium stores in the reticulum could be sufficient to trigger a capacitive entry of calcium, but this has been shown in MCF7 cells for which a protein of the golgi apparatus, called SPCA2 (Secretary Pathway Calcium ATPase 2) , is able to constitutively activate ORAI1 channels (Feng et al., 2010). Such a phenomenon could exist in H23 cells with the ORAI3 channel.

We therefore find that the stimulus causing the capacitive input of calcium is not precisely known in H23 cells and that the frequency of these stimuli is also not known, in particular the capacitive input of calcium take place constantly or at specific times in the life of the cell?

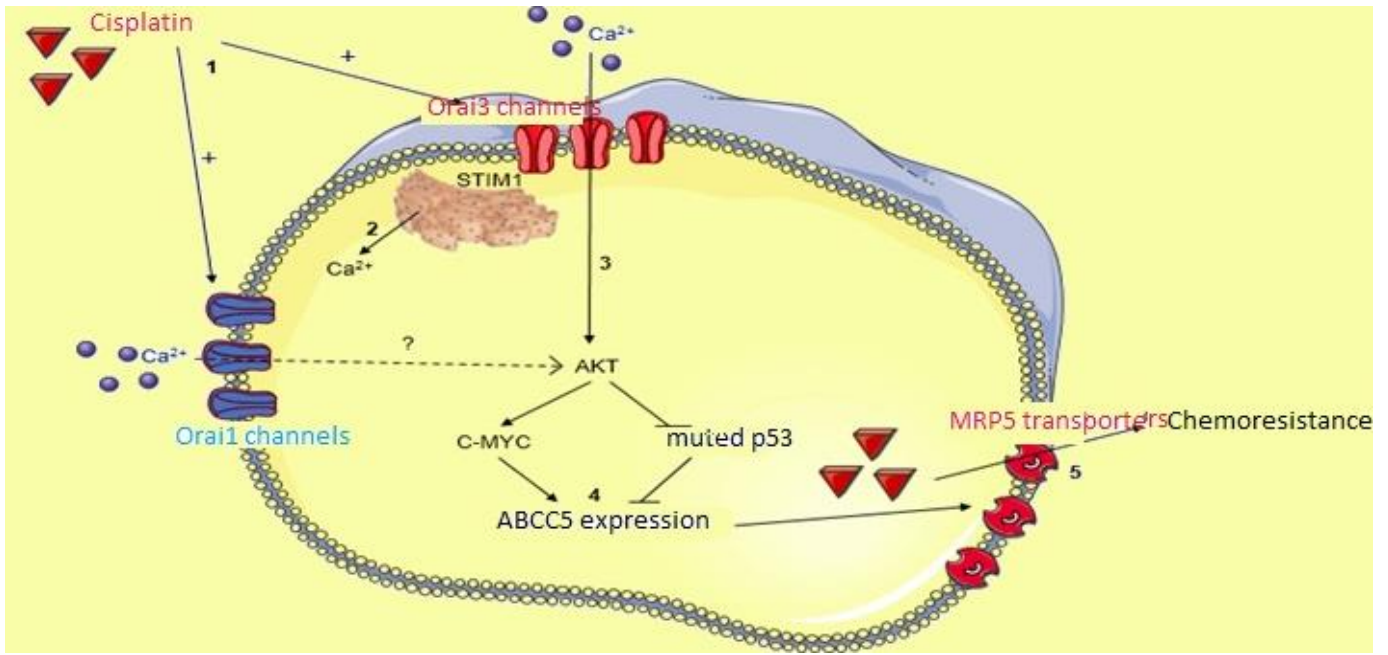

Figure 6: summary diagram of the supposed interactions between the different actors of cisplatin resistance in H23 cells. Administration of cisplatin increases the mRNA levels of the ORAI3 and ORAI1 channels (1), the increase in functionality remains to be determined. In (2) the emptying of calcium stores in the reticulum, at certain times in the life of the cell, triggers the oligomerization of STIM1 which allows the opening of the SOC type channels ORAI3 (and ORAI1, not shown here). The entry of calcium through these channels would allow the activation of the PI3K/AKT pathway as well as potentially other pathways (3). The link between calcium entry and this activation for ORAI1 channels is uncertain. As the AKT pathway is more activated, downstream AKT targets like C-MYC and P53 will be further modulated, C-MYC will be activated and P53 repressed. C-MYC potentially playing a role in the expression of the ABCC5 gene in H23 cells, the activation of the AKT pathway would be associated with an increase in the level of functional MRP5 transporters. The mutated form of P53 in H23 cells would, on the contrary, have an antagonistic role in suppressing the expression of the ABCC5 gene and the inhibition of P53 by AKT would contribute to increasing the levels of functional MRP5 (4). Finally, this increase in the level of functional MRP5 would be associated with an increase in the efflux of cisplatin, reducing the intracellular concentrations of the anti-cancer agent and therefore reducing its effectiveness (5). The mutated form of P53 in H23 cells would, on the contrary, have an antagonistic role in suppressing the expression of the ABCC5 gene and the inhibition of P53 by AKT would contribute to increasing the levels of functional MRP5 (4). Finally, this increase in the level of functional MRP5 would be associated with an increase in the efflux of cisplatin, reducing the intracellular concentrations of the anti-cancer agent and therefore reducing its effectiveness (5). The mutated form of P53 in H23 cells would, on the contrary, have an antagonistic role in suppressing the expression of the ABCC5 gene and the inhibition of P53 by AKT would contribute to increasing the levels of functional MRP5 (4). Finally, this increase in the level of functional MRP5 would be associated with an increase in the efflux of cisplatin, reducing the intracellular concentrations of the anti-cancer agent and therefore reducing its effectiveness (5).

**Conclusion and perspectives**

It then appears from all of these experimental results that the ORAI3 channel seems to play a role in resistance to cisplatin in H23 cells in vitro. Its functional invalidation decreases resistance to cisplatin in H23 cells. This decrease in resistance is correlated with a decrease in the capacitive influx of calcium as well as a decrease in the mRNA levels of the ORAI3 channel, STIM1 and MRP5. On the contrary, cisplatin, in the presence of the functional ORAI3 channel, increases the capacitive entry of calcium as well as the mRNA levels of the ORAI3, ORAI1, STIM1 and MRP5 channels compared to cells without cisplatin and with the functional ORAI3 channel. Chronic cisplatin incubation is furthermore coupled with greater resistance to cisplatin than during acute incubation. This greater resistance is correlated with an increase in mRNA levels of the ORAI3 channel and STIM1 after one month. Taken together, these data suggest that the capacitive calcium input permitted by the ORAI3 channel activates signaling pathways further downstream, such as the PI3K/AKT pathway or other pathways. Then these pathways activate the expression of genes involved in resistance to cisplatin, such as ABCC5 for example.

To complete this work, it will first be necessary to increase the number of samples in order to validate the results already obtained.

Cisplatin induces a reduction in cell viability in H23 cells, but the MTT test does not provide information on how this viability is reduced. Indeed, a reduction in viability may be due to an increase in mortality and/or to a cytostatic effect. For this, a trypan blue test to assess cell death will be necessary to complete this work. Then the flow cytometry technique can be used to establish what type of mortality it is (apoptosis/necrosis), the trypan blue test not allowing to know it.

In addition to this, the MTT test could also be performed after 72 hours of incubation in cisplatin instead of 48 hours to have the best possible match between the cellular events recorded in calcium imaging after 72 hours of incubation and the resistance determined by the test. MTT.

We have seen that cisplatin increases the capacitive calcium input probably by increasing the expression of the ORAI3 channel. It would be necessary to verify that this increase in expression is reflected at the protein and functional levels via the techniques of western blot and patch-clamp, respectively.

The mRNA levels of the ORAI3, ORAI1, STIM1 and MRP5 channels were estimated in the presence of cisplatin alone or of the invalidated ORAI3 channel alone. These levels should be evaluated in the presence of cisplatin combined with the functional invalidation of the ORAI3 channel, which would allow a comparison with the MTT tests carried out in which the functional invalidation of the ORAI3 channel is associated with different doses of cisplatin.

The effect of the decrease in the level of MRP5 mRNA when the ORAI3 channel is invalidated on the functionality of the MRP5 proteins must imperatively be verified. A western blot will firstly allow to know if the protein levels of MRP5 are reduced. Then a functional test should make it possible to highlight a lower efflux of cisplatin by MRP5. Cisplatin efflux tests could then be implemented in the longer term in this study. In the presence and in the absence of the ORAI3 channel, the extracellular cisplatin, therefore expelled from the cell by MRP5, will be measured by means of mass spectrometry. In the absence of the functional ORAI3 channel, the extracellular cisplatin should therefore be present in a lower quantity.

Calcium imaging was performed only on cells incubated in cisplatin for 72 hours, it would then be necessary to carry out this same experiment on cells chronically incubated in cisplatin in order to see if the capacitive entry of calcium is greater than that measured for cells acutely incubated in cisplatin.

The functional invalidation of the ORAI3 channel should also be performed on cells chronically incubated in cisplatin in order to see if they can be resensitized to the anti-cancer agent. To evaluate this resensitization the same techniques would be used: MTT test with determination of the IC50, calcium imaging with determination of the capacitive calcium entry and real-time quantitative RT-PCR with determination of the mRNA level of the different actors studied here. .

The signaling pathways linking capacitive calcium entry to the transcription of these actors need to be elucidated, in particular there would be a probable involvement of the AKT pathway.

The nature of the stimuli allowing the capacitive entry of calcium must be clearly defined in the H23 since this entry seems to play a primordial role in the process of resistance dependent on the ORAI3 channels.

The real-time quantitative RT-PCR carried out after one month on the chronically incubated cells would benefit from being carried out at different times of exposure to cisplatin and not only after one month. In particular to see if there is an increase in mRNA levels of ORAI1 and MRP5 over time compared to acutely incubated cells. Linking these possible increases to calcium signals (capacitive calcium input) corresponding over time would make it possible to establish a temporal link between the calcium signal and the increase at the transcriptional level of the expression of actors involved in cisplatin resistance. This would make it possible to better describe the situation encountered during a course of cisplatin.

Other lines could be studied in the medium term, such as A549 cells, the scientific documentation of which is already well supported in relation to H23 cells. Molecules like oxaliplatin could also be studied in addition to cisplatin. This last molecule constitutes a future alternative to cisplatin for patients with non-small cell lung cancer.

This work would then be integrated in the longer term into a more applied vision via the study of primary and in vivo cultures in murine models of xenografts by evaluating tumor regression following the application of a pharmacological inhibitor of Orai3 channels such as gadolinium.

**References**

1. Abolfazl, A., Ravi, N., Elisa, G. and Godefridus, JP (2016) Role of Akt signaling in resistance to DNA-targeted therapy. World J Clin Oncol. 352-369

2. Anne-Sophie, Ay., Nazim, B., Henri, S., Ahmed A. and Halima, O. (2013) Orai3 Constitutes a Native Store-Operated Calcium Entry That Regulates Non Small Cell Lung Adenocarcinoma Cell Proliferation. PLoS One.

3. Bhaskar, PT. and Hay, N. (2007) The two TORCs and Akt. DevCell. 487-502

4. Dexing, F., David, H., Yanhua, Z., Yan, X., Jill, M., Heinz, N., Gordon, BM, Ryuji, K., Tony, H. and Zhimin, Lu. (2007 ) Phosphorylation of β-Catenin by AKT Promotes β-Catenin Transcriptional Activity. J Biol Chem. 11221-11229

5. Faouzi, M., Kischel, P., Hague, F., Ahmed, A., Benzerdjeb, N., Henri, S., Penner, R. and Halima, O. (2013) ORAI3 silencing alters cell proliferation and cell cycle progression via c-myc pathway in breast cancer cells. Biochim Biophys Acta. 752-60

6. Feng, M., Grice, DM., Faddy, HM., Nguyen, N., Leitch, S., Wang, Y., Muend, S., Kenny, PA., Sukumar, S., Roberts-Thomson, SJ ., Monteith, GR. and Rao, R. (2010) Store-independent activation of Orai1 by SPCA2 in mammary tumors. Cell. 84-98

7. Fraser, M. Bai, T. and Tsang, BK. (2008) ActPromotes cisplatin resistance in human ovarian cancer cells through inhibition of p53 phosphorylation and nuclear function. Int J Cancer. 534-46

8. Gregory, MA., Qi, Y. and Hann, SR. (2003) Phosphorylation by glycogen synthase kinase-3 controls c-myc proteolysis and subnuclear localization. J Biol Chem.

9. Guminski, AD., Balleine, RL., Chiew, YE., Webster, LR., Tapner, M., Farrell, GC., Harnett, PR. and Defazio, A. (2006). MRP2 (ABCC2) and cisplatin sensitivity in hepatocytes and human ovarian carcinoma. Gynecol Oncol. 100, 239-285.

10. Lelj-Garolla, B., Kumano, M., Beraldi, E., Nappi, L., Rocchi, P., Ionescu, DN., Fazli, L., Zoubeidi, A. and Gleave, ME. (2015) Hsp27 Inhibition with OGX-427 Sensitizes Non-Small Cell Lung Cancer Cells to Erlotinib and Chemotherapy. Mol Cancer Ther. 1107-16

11. Lin, W., Xingxiang, P., Qianzhi, W., Jun, C., Fang, X., Li, X. and Kang, L. (2016) miR-96 induces cisplatin chemoresistance in non-small cell lung cancer cells by downregulating SAMD9. Oncol Lett. 945-952

12. Lindsey, DM and David, BD (2001) A phosphatidylinositol 3-kinase/Akt pathway promotes translocation of Mdm2 from the cytoplasm to the nucleus. Proc Natl Acad Sci USA. 11598-11603

13. Michael, RG, Michael, PS, Lorna, S., May, D., Richard, AR, Linda, M., Vincent, D. and Danielle, LK (1999) The B cell antigen receptor activates the Akt (protein kinase B )/Glycogen synthase kinase-3 signaling pathway via phosphatidylinositol 3-kinase. J. Immunol. 163, 1894-1905.

14. Naruji, K., Arata, N., Tohru, H., Koji, U., Tadahiko, E., Tao-Sheng, L. and Kimikazu, H. (2015) The c-MYC-ABCB5 axis plays a pivotal role in 5-fluorouracil resistance in human colon cancer cells. J Cell Mol Med. 1569-1581

15. Selvaraj, S., Sun, Y., Sukumaran, P. and Singh, BB. (2016) Resveratrol activates autophagic cell death in prostate cancer cells via downregulation of STIM1 and the mTOR pathway. Mol Carcinog. 818-31

16. Shiuh-Dih, C., Thomas, P., Jianlin, Gong. and Stuart KC (2012) mTOR Is Essential for the Proteotoxic Stress Response, HSF1 Activation and Heat Shock Protein Synthesis. PLoS One.

17. Wakamatsu, T., Nakahashi, Y., Hachimine, D., Seki, T. and Okazaki, K. (2007). The combination of glycyrrhizin and lamivudine can reverse the cisplatin resistance in hepatocellular carcinoma cells through inhibition of multidrug resistance-associated proteins. Int J Oncol 1465-72

18. Wang, Q. and Beck, WT. (1998) Transcriptional suppression of multidrug resistance-associated protein (MRP) gene expression by wild-type p53. Cancer Res. 5762-9

19. Wolfgang, H., Ralf, F., Martina, S., Matthias, L. and Ralf, J. (2011). Membrane Drug Transporters and Chemoresistance in Human Pancreatic Carcinoma. Cancers (Basel) 106-125

20. Xu, Y., Diao, Y., Qi, S., Pan, X., Wang, Q., Xin, Y., Cao, X., Ruan, J., Zhao, Z., Luo, L., Liu, C. and Yin, Z. (2013) Phosphorylated Hsp27 activates ATM-dependent p53 signaling and mediates the resistance of MCF-7 cells to doxorubicin-induced apoptosis. Cell Signal. 1176-85

21. Xu, ZP., Devillier, P., Xu, GN., Qi, H., Zhu, L., Zhou, W., Hou, LN., Tang, YB., Yang, K., Yu, ZH., Chen, HZ. and Cui, YY. (2013) TNF-α-induced CXCL8 production by A549 cells: Involvement of the non-neuronal cholinergic system. Pharmacol Res. 16-23

22. Zhang, Y., Bao, C., Mu, Q., Chen, J., Wang, J., Mi, Y., Sayari, AJ., Chen, Y. and Guo, M. (2016) Reversal of cisplatin resistance by inhibiting PI3K/Akt signal pathway in human lung cancer cells. Neoplasma. 362-70

23. Zhao, Q., Gu, X., Zhang, C., Lu, Q., Chen, H. and Xu, L. (2015) Blocking M2 muscarinic receptor signaling inhibits tumor growth and reverses epithelial-mesenchymal transition (EMT) in non-small cell lung cancer (NSCLC). Cancer Biol Ther. 634-43